\documentclass[aps,pra,reprint,showpacs,floatfix]{revtex4-1}

\usepackage{amsmath}
\usepackage{graphicx}
\usepackage[
breaklinks=true,
pdfauthor={Michael Klaiber},
pdftitle={Multiphoton excitation}
]{hyperref}
\usepackage{color}
\def\sout{\bgroup\markoverwith
{\textcolor{red}{\rule[0.5ex]{2pt}{0.5pt}}}\ULon}

\graphicspath{{figures/}}

\renewcommand*{\vec}[1]{\boldsymbol{#1}}

\newcommand{\eref}[1]{Eq.~(\ref{#1})}
\newcommand{\bra}[1]{\langle\,{#1}\, |}
\newcommand{\ket}[1]{|\,{#1}\,\rangle}
\newcommand{\braket}[2]{\mbox{$\langle\,{#1}\, | \,{#2}\,\rangle$}}
\usepackage{ulem} 
\usepackage{lineno}

\begin{document}
\title{The cross-over from tunnelling to multiphoton ionization of atoms}

\author{Michael Klaiber}
\email{klaiber@mpi-hd.mpg.de}
\affiliation{Max-Planck-Institut
f\"ur Kernphysik, Saupfercheckweg 1, 69117 Heidelberg, Germany}
\author{John S. Briggs}
\email{briggs@uni-freiburg.de}
\affiliation{Institute of Physics, University of Freiburg, Germany}

\date{\today}

\begin{abstract}
We present a theory illuminating the cross-over from strong-field tunnelling ionization to weak-field multiphoton ionization
in the interaction of a classical laser field with a hydrogen atom. A simple formula is derived in which the ionization amplitude appears as a product of two separate amplitudes. The first describes the initial polarization of the atom by virtual multiphoton absorption and the second the subsequent tunnelling out of the polarized atom. Tunnelling directly from the ground state and multiphoton absorption without tunnelling appear naturally as the limits of the theory.

\end{abstract}


\maketitle

\section{Introduction}
Traditionally the interaction of a strong laser field with an atom which leads to ionization is considered to occur by two contrasting mechanisms, as limiting cases of the so-called Keldysh parameter $\gamma = \sqrt{\frac{I_p}{(2U_p)}}$ where $I_p$ is the atom ionization potential and $U_p$ is the laser ponderomotive potential.  The Keldysh parameter can be represented in atomic units also as  $\gamma =  \omega(2I_p)^{1/2}/E_0$, where the classical laser electric field has frequency $\omega$ and strength $E_0$. For $\gamma < 1$ or low frequency and high intensity, ionization proceeds by tunnelling out of the atomic potential under the influence of the potential supplied by the field. When $\gamma > 1$, for high frequency and relatively low intensity, then direct absorption of  several photons occurs giving rise to ionization and possibly "above-threshold" ionization (ATI) due to absorption of further photons by electrons already in the continuum.

The earliest theoretical treatments were based on  the "strong-field approximation" (SFA) by Keldysh \cite{Keldysh}, Faisal \cite{Faisal} and Reiss \cite{Reiss}, known as the KFR approach. More recent developments e.g.\cite{IMS},\cite{Becker}, \cite{Popruz} contain extensions of the SFA and in certain cases fully numerical calculations are now available \cite {Smirnova},\cite{Christoph}, \cite{JMR}. 

The SFA and its extensions have been very successful  in describing the ionization of atoms by laser fields. It involves the approximation of the exact T-matrix element for the ionizing transition by a matrix element of the form
\begin{equation}
\label{SFA2}
f = \int_{-\infty}^{\infty}  \bra{\phi_f(t')}V_F(t')\ket{\phi_i(t')} ~dt',
\end{equation}
where $\ket{\phi_i}$ is the initial eigenstate of the atom alone, $\ket{\phi_f}$ is the continuum eigenstate of the laser field alone and
$V_F$ is the interaction of atomic electron with laser field.
From the outset it was shown that both limits are contained in this theory. This is plausible if one considers that in a Floquet picture 
$\ket{\phi_f}$ contains the electron coupled to any number of photons, so that multiphoton ionization is described. Additionally, in the opposite limit of $\gamma < 1$, a semi-classical description of $\ket{\phi_f}$ in the classical laser field corresponds to a tunnelling interpretation. Nevertheless, a clear physical picture of the intermediate region and the transition between the two limits does not emerge.

 Recently there has been renewed interest in developing the theory to explain more physically how the transition from direct tunnelling out of the ground state to the seemingly quite different mechanism of multi-photon absorption occurs. A first step in this direction was made by Klaiber et.al.  \cite{Klaiber1} who recognized that during tunnelling the electron may absorb energy from the field to emerge at the tunnel exit with more energy than when it started to tunnel. This "non-adiabatic" energy absorption was treated by classical mechanics. Clearly however, as will be explained in more detail below, the energy gain can also be thought of as the absorption of photons during the tunnelling process.
 
In a recent letter \cite{russian} a numerical study of ionization in extremely strong laser fields was reported. The results were interpreted by pointing out that tunnelling may not only occur directly out of the atomic ground state (the hydrogen atom was used as example). Rather, excitation of the atom to higher bound states may occur followed by tunnelling. Indeed, since the binding energy is then much reduced, as in the non-adiabatic case, tunnelling is more efficacious from such excited states. In fact the picture presented corresponds to "over-the-barrier" ionization out of excited states without the need for tunnelling.  Although plausible, we feel that such a picture is not quite correct. The authors considered that the {\it{real}} time-independent hydrogen eigenstates must first be populated by multi-photon absorption. Since for the case of hydrogen the first excited $N=2$ manifold lies $10.4eV$ above the ground state, a significant multiphoton transition is required to populate these states. Higher manifolds were not considered.

Here we put forward an alternative picture in which the dominant role is played not by real eigenstates but by virtual "off the energy shell" states of the hydrogen atom. For infra-red and visible photons these energy states lie far below the $N = 2$ excited manifold but nevertheless tunnelling can occur from them. We will show that the picture of virtual absorption allows a simple description of the smooth transition from tunnelling to multiphoton ionization to be given. We make clear from the outset that our aim is not to develop a theory with which to confront specific experimental data. There one cannot compete with fully numerical methods. Rather it is to expose the physical mechanism of laser-atom interaction and to explain in a simple picture how the cross-over from tunnelling to multiphoton ionization arises.

 When an atom is subject to an electric field whose frequency is not resonant with transition to an eigenstate, there is an interaction and distortion which is usually referred to classically as polarization of the electron cloud. In the quantized photon picture this is ascribed to the virtual absorption of photons. After each photon absorption there is a changed wavefunction and since the energy is higher, this wavefunction usually extends to larger distance. That is, as each photon is absorbed virtually, the atom "swells" in extent. Clearly, ionization by real photon absorption or by tunnelling, can occur readily from such extended and weakly-bound virtual states.
 A calculation of this process of virtual excitation followed by tunnelling is the subject of this paper.
 
 Already in 1988 the virtual multiphoton off-shell atomic states were used to provide the first explanation of electron angular distributions in ATI of the hydrogen atom \cite{Kracke} by comparison with experiments of Feldmann et.al. \cite{Feld}.
 Unfortunately, although studied extensively in unpublished work of Kracke \cite{Guido} and used in ion-atom collisions \cite {Marxer}, no further discussion of the nature and properties of such wavefunctions seems to have been published. 

 The plan of the paper is as follows. In section II we present a critique of the standard scattering theory used to calculate the transition amplitude (T-matrix element) to continuum states. We  show that the transition amplitude to a continuum state can be represented in an intuitively appealing way in that it appears as a direct product of the amplitude for virtual $n$-photon absorption and the probability for subsequent tunnelling from this virtual excited state. The results of calculations for the realistic case of the three-dimensional hydrogen atom are presented in section III.  The virtual absorption wavefunction is calculated numerically by iteration of the inhomogeneous Schr\"odinger equation and  for the tunnelling wavefunction the quasi-static approximation is employed using a separation in parabolic coordinates. The results indeed exhibit a smooth and continuous transition from optimum tunnelling directly from the ground state, to dominant multiphoton ionization, as the Keldysh parameter is varied. Throughout we use atomic units in which the electron charge, the electron mass and $\hbar$ are equal to unity.

\section{The basic equations}
\subsection{The strong-field approximation}
The dynamics of ionization of an initially-bound electron in a strong laser field is essentially decided by the competition between two electric fields; that of the parent nucleus and that of the external laser. As such there is great similarity with the theory of electron exchange in ion-atom collisions where the two competing fields are those of the two nuclei and involve two frames of reference, the laboratory frame of the parent nucleus and the moving frame of the incident nucleus. Indeed "over the barrier" ionization was first formulated for the ion-atom problem. This analogy will emerge also in the formulas presented here and perhaps casts a new light on the SFA.

We consider a total Hamiltonian
\begin{equation}
H(t) = H_i + V_F(t) = K + V  + V_F(t) = H_f + V,
\end{equation}
where $K$ is the electron kinetic energy operator, $V$ is the nuclear potential and  $V_F(t)$ is the interaction between electron and laser field (considered to be a classical field).
The electron wavefunction at time $t$ is given by solution of the equation
\begin{equation}
H(t)\Psi(t) = i \frac{\partial \Psi(t)}{\partial t}.
\end{equation}
The transition probability amplitude from an initial to a final state at time $t$ can be expressed in two equivalent post and prior forms, i.e.
\begin{equation}
\label{amplitude}
f(t) = \braket{\phi_f(t)}{\Psi^+_i(t)} = \braket{\Psi^-_f(t)}{\phi_i(t)}.
\end{equation}
The two exact wavefunctions propagate forward in time with $\Psi^+_i(t) \rightarrow \phi_i$ as $ t \rightarrow -\infty$ and
backwards in time with $\Psi^-_f(t) \rightarrow \phi_f$ as $ t \rightarrow \infty$ respectively. If one considers that $\phi_i$ is an eigenstate of $H_i$ and $\phi_f$ is an eigenstate of $H_f$, then from the Schr\"odinger equation one can show that
\begin{equation}
\label{post}
f(t) = \int_{-\infty}^t \bra{\phi_f(t')}V\ket{\Psi^+_i(t')} ~dt'
\end{equation}
for the post form or
\begin{equation}
\label{prior}
f(t) = \int^{\infty}_t \bra{\Psi_f^-(t')}V_F(t')\ket{\phi_i(t')} ~dt'
\end{equation}
for the prior form.
These two expressions are exact.\\
 An approximation that has received much attention for ionization is the SFA of \eref{SFA2}. 
In the formalism of re-arrangement given here, one notes that $\phi_f(t)$ is an eigenstate of the (electron + field) Hamiltonian $H_f$.
Hence this is a Volkov state and the SFA is made simply by replacing  $\Psi^+_i$ in \eref{post} by the initial state $\phi_i$. Interestingly, although often termed "non-perturbative" now the SFA appears as the first Born term for re-arrangement of the electron between eigenstates of the two potentials
\begin{equation}
\label{SFA1}
f^{SFA} = \int_{-\infty}^{\infty}  \bra{\phi_f(t')}V\ket{\phi_i(t')} ~dt'.
\end{equation}

As in general re-arrangement scattering \cite{MacD}, one can show that the equivalent first Born approximation putting 
$\Psi_f^- \approx \phi_f$ in the prior form \eref{prior} is identically equal i.e.
\begin{equation}
\label{SFA3}
f^{SFA} = \int_{-\infty}^{\infty}  \bra{\phi_f(t')}V_F(t')\ket{\phi_i(t')} ~dt',
\end{equation}
so that one can use either potential in the first Born SFA re-arrangement matrix element.

 In the length gauge the Volkov state reads (in units with $ e= \hbar = m = 1$ and $c=137$)
\begin{equation}
  \phi_f^V(\vec r,t)=\exp[i(\vec p+\vec A(t)/c].\vec r-i\int^tdt'(\vec p+\vec A(t')/c)^2/2]
\end{equation}
where  $\vec A$ is the vector potential. This is simply the Kramers-Henneberg space-translated plane wave \cite{Faisal} describing the electron stationary in the moving field. The additional exponential energy and momentum factors involving $\vec A$ are identical to the "electron translation factors" appearing on final state wavefunctions in ion-atom electron capture, where the electron is also stationary in the moving field of the ion \cite{BriggsMacek}. This justifies our view of the SFA as a collisional re-arrangement process in first Born approximation.

In approximate evaluations in collision theory, the time integrated forms \eref{post}, \eref{prior} of the transition amplitude are usually preferred to the direct projection forms \eref{amplitude}. Basically this is because, if $\phi_i$ and $\phi_f$ are orthogonal as is usually the case, the forms
 \eref{amplitude} give zero  for the first-order amplitude whereas the integral forms \eref{SFA1} and \eref{SFA3} give a finite result.
 
  By contrast, in numerically accurate propagations of the time-dependent wavefunction it is more direct to use the projection form \eref{amplitude}. This will be the strategy adopted in this paper and it removes a certain ambiguity in the physical interpretation of the SFA when the two equivalent forms
 \eref{SFA1} and \eref{SFA3} are used. In the form \eref{SFA1} one would say that ionization out of the initial state occurs by the electron scattering from its parent nucleus and then accessing the Volkov state, describing either tunnelling or absorption of
 photons depending upon the value of $\gamma$. However, the form \eref{SFA3} would be interpreted as an initial absorption of a
 single photon via $V_F$,  followed by overlap on the same Volkov state. Which physical picture is correct ?
 
In the following we describe ionization in a unified way in that we approximate  $\Psi^+_i$ in \eref{post}  essentially by a {\it{product}}      
 of a state which initially has absorbed virtually a certain number of photons and a semi-classical state describing subsequent tunnelling in the full potential of the nuclear and laser electric fields. This describes a continuous transition from tunnelling to multiphoton regime according as to which element of the product states is dominant.

\subsection{The approximate transition matrix element}

We begin, not with the standard form \eref{post} of the transition amplitude, but with the direct time propagation of \eref{amplitude}
\begin{equation}
f(t) = \braket{\phi_f(t)}{\Psi^+_i(t)} = \bra{\vec p_f}U(t,-\infty)\ket{\phi_i}
\end{equation}
where $ \bra{\vec p_f}$ is the final momentum state of the continuum electron, $\ket{\phi_i}$ is the initial atomic state and 
\begin{equation}
\left(H(t) - i\frac{\partial}{\partial t}\right)U(t,t') = 0
\end{equation}
 is the full time propagator. Our approach will be to approximate the time development as occurring initially, up to a time $t_i$ with the laser field as a perturbation, followed by a propagation in the static field of the laser plus atomic Coulomb potential. That is, we write the full time-development operator as a product,
\begin{equation}
\begin{split}
f(t)& = \bra{\vec p_f}U(t,t_i) U(t_i,-\infty)\ket{\phi_i}\\& = \bra{\vec p_f}U(t,t_i)\ket{\psi(t_i)}
\end{split}
\end{equation}
where $\ket{\psi(t_i)}$ is an  off-shell atomic state with photons absorbed virtually. This describes an initial polarization of the atom by the laser field. In the next section this state will be expanded in states in which a given number $n$ of photons has been absorbed virtually. The operator $U(t,t_i)$ then describes the subsequent tunnelling transition of the electron to a final ionized state. Then we write
\begin{equation}
\begin{split}
& \bra{\vec p_f}U(t,t_i)\ket{\psi(t_i)}\\& = \int d\vec r_f  \braket{\vec p_f}{\vec r_f} \bra{\vec r_f}U(t,t_i)\ket{\psi(t_i)}.
 \end{split}
\end{equation}
Since the propagation through the tunnelling region and beyond as a continuum electron will be described subsequently by a semi-classical wavefunction, we will define ionization probability as given by the probability density $| \bra{\vec{r}_f}U(t,t_i)\ket{\psi(t_i)}|^2$ at a point $\vec r_f$, corresponding to the exit from the tunnelling region. In appendix B it is shown that, as a result of the {\it{imaging theorem}} \cite{JBJF}, this is equal to the transition probability $| \bra{\vec p_f}U(t,t_i)\ket{\psi(t_i)}|^2$.

The matrix element to be calculated can be written as an integral of the product of the transition amplitude to a virtual state multiplied by the tunnelling amplitude i.e.

\begin{equation}
\label{poswf}
\begin{split}
& \Psi^+(\vec r_ft_f) \equiv \bra{\vec r_f}U(t_f,t_i)\ket{\psi(t_i)}\\& = \int  \bra{\vec r_f}U(t_f,t_i) \ket{\vec r_i}\braket{\vec r_i}{\psi(t_i)}~ d\vec{r}_i
\\& = \int  K(\vec r_f t_f,\vec r_i t_i)\, \psi(\vec r_i t_i)~d\vec{r}_i
 \end{split}
\end{equation}
where we have introduced the kernel $K(\vec r_f t_f,\vec r_i t_i)$.

\section{The off-shell wavefunctions}

We consider ionization of a hydrogen atom i.e. we take $H_i \equiv H_0 = K + V$ where  $V(r)=-\kappa/r$, the ionization potential $I_p= \kappa^2/2$ and nuclear charge $\kappa=1$.
The initial bound  $1s$ ground state is
\begin{eqnarray}
 {\psi}_0(r)=\sqrt{\frac{\kappa^3}{\pi}}\exp[-\kappa r].
\end{eqnarray}

\begin{figure} 
    \begin{center}
      \includegraphics[width=0.4\textwidth]{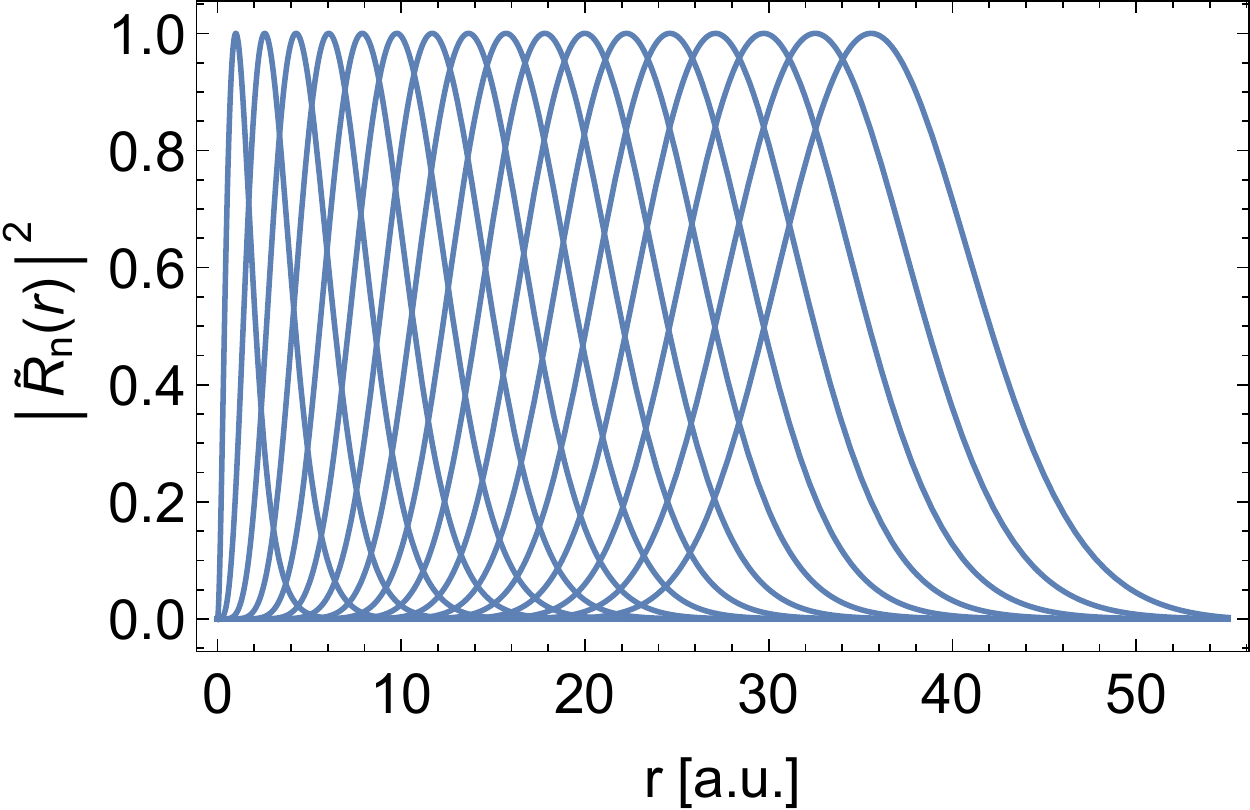}
	\caption{The off-shell Coulomb wavefunctions for $n$-photon virtual absorption with $\omega = 0.025$ a.u.. From left to right $n = 0$ to $16$.}
\label{ofwfC}
    \end{center}
  \end{figure}

The first task is to calculate the virtual  state $\ket{\psi(t_i)}$. We approximate the
exact state by its lowest-order perturbation result. Hence, for this part of the ionization process the laser field is taken to be effectively
a c.w. pulse. Then,
the hydrogen atom is driven by a periodic circularly polarized laser with electric field $\mathbf{F}(t)= E_0 \hat{\bf x} \cos[\omega t] + E_0 \hat{\bf y} \sin[\omega t]$. The time-dependent Schr\"odinger-equation then reads
\begin{eqnarray}
H_0\ket{\psi(t)}-\mathbf{r}\cdot\mathbf{F}(t)\ket{\psi(t)} - i \frac{\partial\ket{\psi(t)}}{\partial t} = 0
\label{TDSE}
\end{eqnarray}
Since the laser field is periodic the state vector can be expanded in a Floquet Fourier series
\begin{eqnarray}
\ket{\psi(t)}=  \exp[i I_p t]\sum_j\ket{\psi_j}\exp[-ij\omega t]
\end{eqnarray}
Inserting this expression into the Schr\"odinger equation, multiplying by $\exp[in\omega t]$ and integrating over all time yields
\begin{eqnarray}
\begin{split}
[H_0 + (I_p - n\omega)]\ket{\psi_n} =& \frac{E_0}{2}[(x + iy)\ket{\psi_{n-1}}\\&
 + (x - iy)\ket{\psi_{n+1}}]
\end{split}
\end{eqnarray}
This equation describes the population of the state $\ket{\psi_n}$ by absorption or emission of a photon from neighboring states. Since we consider the initial state as the ground state, in accordance with perturbation theory we retain only the lower state in the inhomogeneous term to give
\begin{eqnarray}
\label{offshellwfn}
\begin{split}
[H_0 + (I_p - n\omega)]\ket{\psi_n} =&  \frac{E_0}{2}(x + iy)\ket{\psi_{n-1}}.
\end{split}
\end{eqnarray}
This is the inhomogeneous equation for the off-shell Coulomb wavefunctions. In \cite{Kracke} it was solved iteratively in a numerical procedure for the absorption of up to nine photons to calculate angular distribution of ATI continuum electrons. Here we restrict discussion to virtual states which are still bound. In appendix A we show how the inhomogeneous equation for the radial wavefunction is derived and solved. The method goes back to Dalgarno and Lewis  in 1955 \cite{Dal-Lew} and has been used often in early work on multiphoton ionization e.g. in \cite{Mizuno}, \cite{Crance}, \cite{Starace}. The results are shown on Fig.\ref{ofwfC}. for the radial density $|\tilde R_n(r)|^2$ as a function of distance $r$
from the nucleus. Since the absolute magnitude of the virtual wavefunctions decrease strongly with $n$, we have normalized each magnitude to unity by defining $\tilde R_n(r) \equiv R_n(r)/R_n(r_n)$, where $r_n$ is the position of the wavefunction maximum. For circular polarization the orbital angular momentum quantum numbers $(l,m)$ are simply $l = m = n$.

The absorption of multiple photons is usually depicted as a vertical process in the atomic potential  but the main feature of the off-shell wavefunctions shown in Fig.\ref{ofwfC} is that, as the energy  and angular momentum of the electron increases, the wavefunction has its maximum at larger and larger $r$ values. In Fig.\ref{ofwfC}, to illustrate clearly the shift of the wavefunction from the nucleus, the modulus squared of each wavefunction for successive photon absorption has been normalized by dividing by its maximum value . The actual magnitude of the wavefunction decreases with each iteration due to the $E_0$ factor. Of course, the shift of the wavefunction to larger distances as binding energy decreases is also a feature of the on-shell eigenstates of the hydrogen atom. The effect is amplified here by the dipole operator in the inhomogenous term. The important point for subsequent tunnelling is that this virtual wavefunction has significant amplitude in the tunnelling region of the combined atomic and laser electric field potential. As shown in the next section, this leads to enhanced tunnelling out of virtually-excited states compared to that from the ground state..

It is interesting to compare the energy gain by virtual  photon absorption treated here, with the energy gain calculated in the ``non-adiabatic" tunnelling picture of \cite{Klaiber1}. Since the former is calculated in quantum picture and the latter in classical mechanics, the quantities to be compared are somewhat arbitrary. However, to be precise, we plot the effective total radial energy of the quantum case against the total energy of the classical case, both evaluated in the tunnelling direction $x$. That is, we plot $-(\kappa_n^2 + n(n+1)/r^2)/2$ (see \eref{Aradial}) at the maximum value of the wavefunction against the energy $(\dot r^2/2 -rE_0 - \kappa/r)$ for the classical energy gain \cite{Klaiber1}. This comparison is shown in Fig.2. 

In the cases $\gamma= 0.66$ and $\gamma = 2.0$, Fig.2 a and b, with low frequency $\omega = 0.025$ a.u.
 there is 
reasonable quantitative agreement but good qualitative agreement. For $\gamma = 0.66$, the frequency implies that the energy increase, due to five photons absorbed, is small on the energy scale shown. However, the quantum calculation shows the wavefunction penetrating 
into the tunnelling region as photons are absorbed. The same is true for the multiphoton ionization regime $\gamma= 6.0$ shown in  Fig.2c.
Here one observes 16 photons absorbed virtually with the energy increasing as a function of position in almost exactly the same way as in the classical calculation. Of course many photons corresponds to the classical limit but the close agreement of the two estimates of energy versus position is quite noteworthy.

\begin{figure} 
    \begin{center}
      \includegraphics[width=0.4\textwidth]{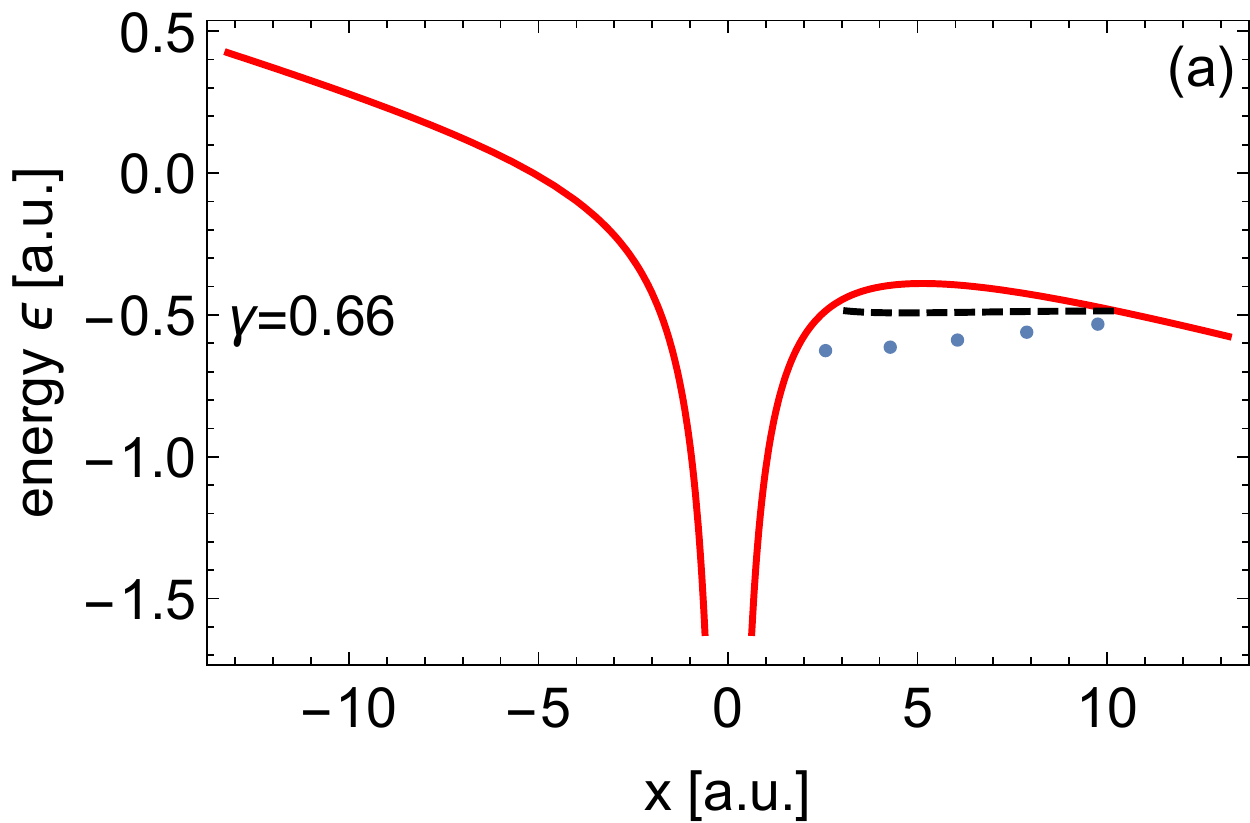}
       \includegraphics[width=0.4\textwidth]{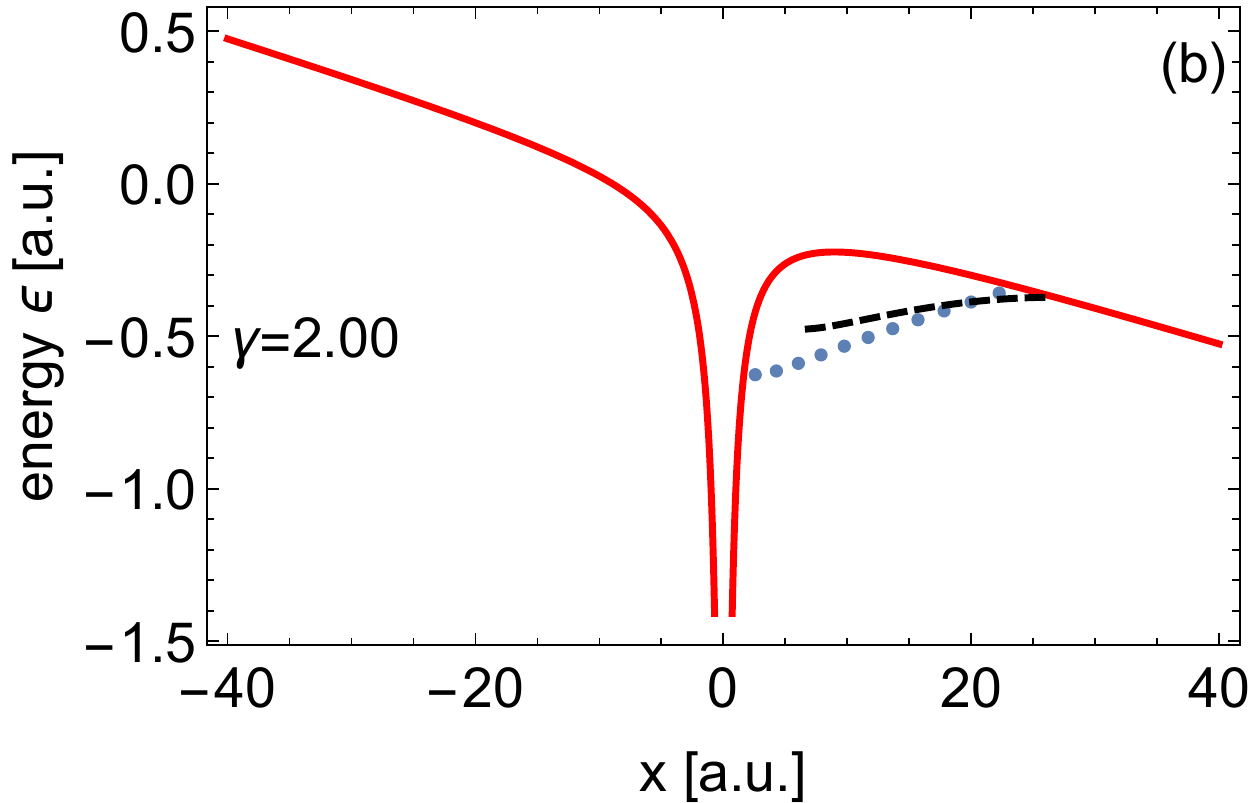}
        \includegraphics[width=0.4\textwidth]{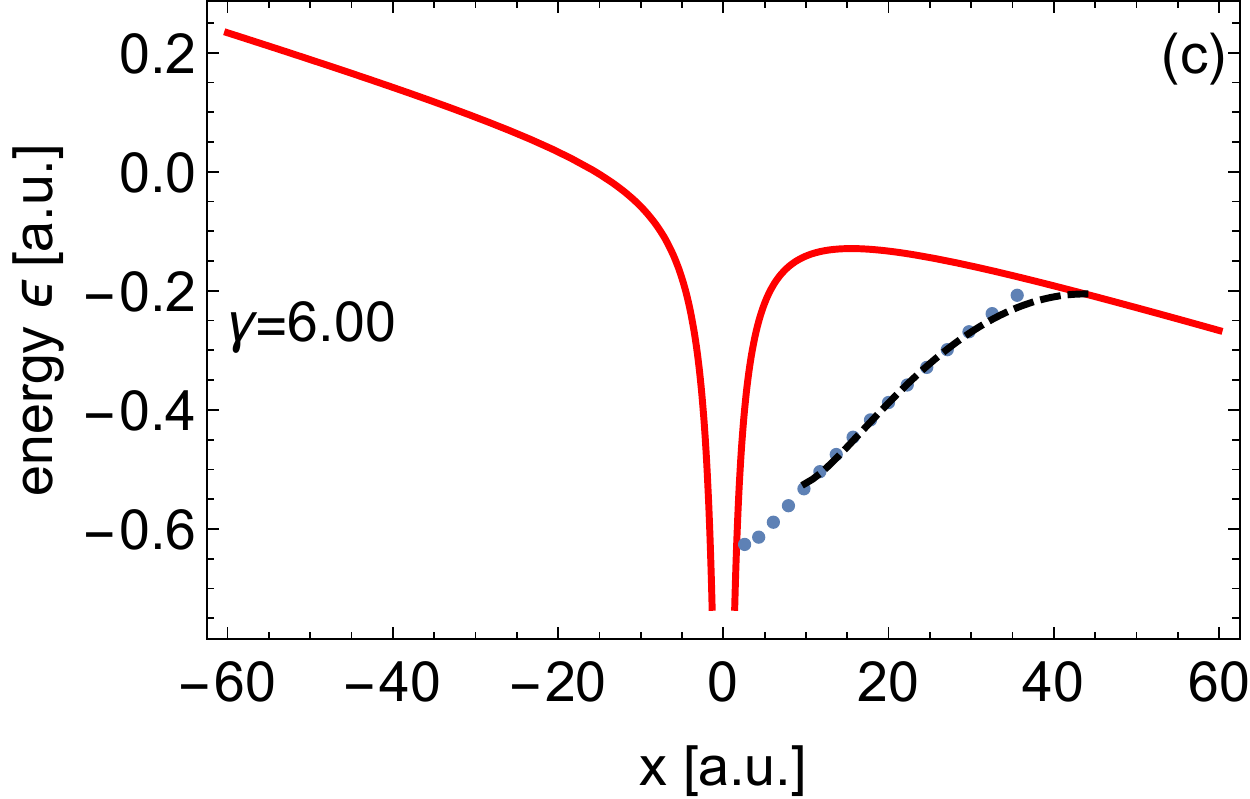}
\label{energy}
\caption{The total potential in the $x$ direction. The dots indicate the position of the wavefunction maximum as a function of energy for increasing number of photons. The dashed line is the classical energy gain.}
    \end{center}
  \end{figure}

\section{The tunnelling wavefunction}
The second task is the calculation of the tunnelling probabilities for different starting values of $\mathbf{r}_i$ in \eref{diffprob} and
integration over $\mathbf{r}_i$ to obtain total ionization probability. \eref{diffprob} we calculate the wavefunction at tunnel exit
\begin{equation}
\label{intri}
\begin{split}
 \Psi^+(\vec r_f)& = \int   K(\vec r_f, \vec r_i )\, \psi_n(\vec r_i )~d\vec{r}_i \\&
 \end{split}
\end{equation}
 The tunnelling wavefunction i.e. the kernel as a function of $\vec r_i$, with the (static) field of strength $E_0$ in the $x$ direction,  satisfies the Schr\"odinger equation
\begin{equation}
\left(-\frac{\Delta}{2} - \frac{1}{r} -xE_0\right) K = -\frac{\kappa_n^2}{2} K
\end{equation}
where $\kappa_n^2 \equiv 2(I_p - n\omega)$.
 Following \cite{LL} and \cite{DarkoLars}
the tunnelling wavefunction is calculated in parabolic coordinates $(\eta,\xi,\phi)$ with $x= (\eta-\xi)/2,y=\sqrt{\eta\xi}\cos{ \phi},z = \sqrt{\eta\xi}\sin{ \phi}$ since the equation separates in these coordinates. The tunnelling is described by the $\eta$ equation and, as shown in detail in \cite{BisLars}, can taken to the lowest order in $\hbar$ i.e. the WKB solution. This gives the semi-classical tunnelling wavefunction. Of course the virtual wavefunction $\psi_n(\vec r_i )$ in \eref{intri} is calculated in spherical coordinates with the $z$ axis perpendicular to the plane of polarization. Then, for circular polarization, the state with angular momentum $l = n$ is populated and with the highest $m = n$ value. The electron density in the excited state is aligned in the $xy$ plane and correspondingly we take $m = 0$ only with respect to the parabolic $\phi$ dependence, which gives optimum tunnelling  \cite{BisLars}. With these approximations the function $K(\vec r_f, \vec r_i)$ can be calculated and the integral over $\vec r_i$ in \eref{intri} performed numerically.
Note that, in parabolic coordinates, the integrand is exactly of the form considered by Landau and Lifshitz \cite{LL} for ionization from the ground state. In their calculation they simply assumed a particular starting point $\vec r_i$. Here we have performed the integral over all $\vec r_i$.

The final  ionization probability is a product of the two competing processes of multi-photon absorption and under-the-barrier tunnelling. 
In perturbation theory, the $n$-photon wavefunction $\psi_n(\vec r_i t_i)$ contains a time-dependent phase factor $\exp{[i(I_p - n\omega)t_i]}$. Hence we will treat each $n$-photon state separately corresponding to a different final energy. Also, to make the calculation of tunnelling probability tractable, as in \cite{Klaiber1}, we will describe tunnelling in the static electric field at a time corresponding to the maximum of the field strength. Then the differential ionization probability out of a state with $n$ photons absorbed virtually is
time independent 

\begin{equation}
\label{diffprob1}
\frac{dP_n}{d \vec r_f} = \left|\Psi^+_n(\vec r_ft_f)\right|^2 =  \left| \int   K(\vec r_f, \vec r_i )\, \psi_n(\vec r_i )~d\vec{r}_i \right|^2.
\end{equation}

The $n$-photon absorption probability decreases as $E_0^{2n}$ whereas the tunnelling probability increases exponentially in $(I_p - n\omega)$.  Below, we consider a fixed frequency, low enough to justify the quasi-static tunnelling approximation but requiring many photons to be absorbed to reach the ionization threshold. Varying $\gamma$ then corresponds to varying field strength.The maximum value of field strength determines the height of the potential barrier for tunnelling and so has a decisive effect on the tunnelling probability. The competition between the probability to access a state by photon absorption and the probability to tunnel out from that state decides the dominant mode of photo-ionization.
 The results illustrate this influence of the laser field strength on the ionization mechanism and are presented in Fig.\ref{integratedprob}. 

\section{The ionization probabilities}
\begin{figure} 
    \begin{center}
      \includegraphics[width=0.4\textwidth]{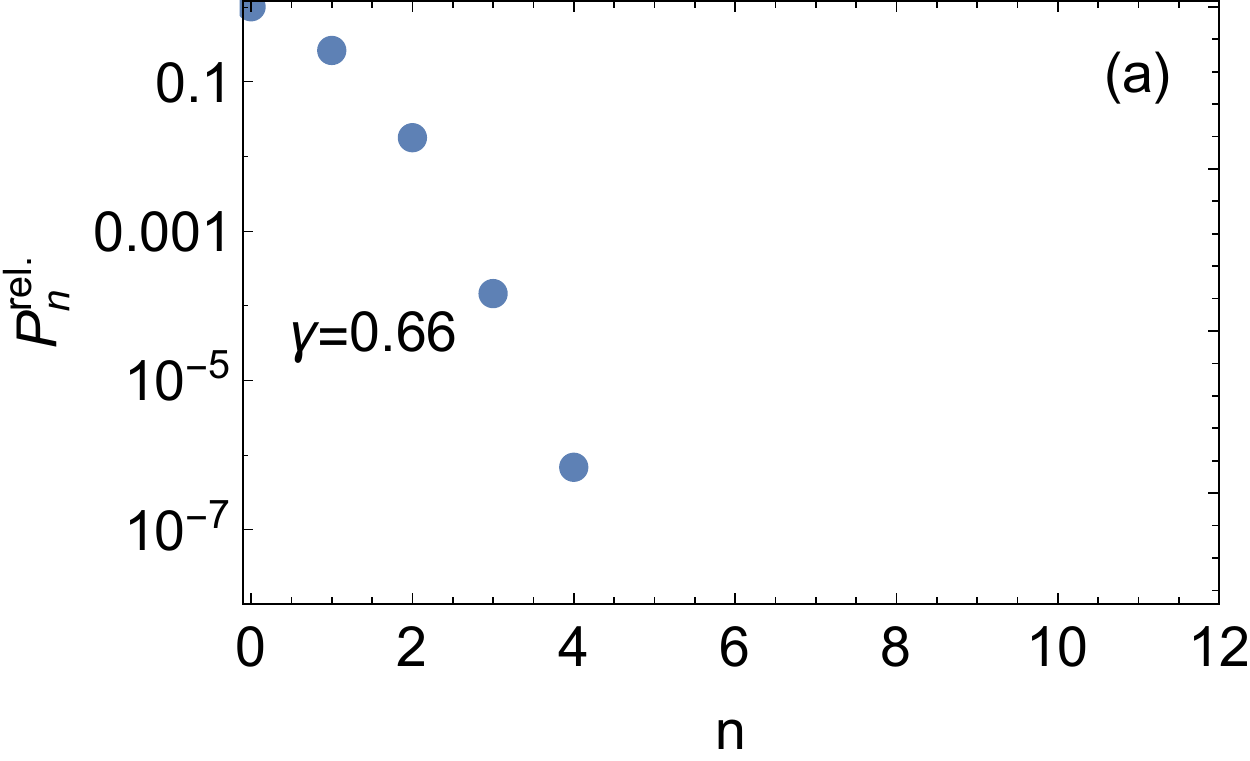}
        \includegraphics[width=0.4\textwidth]{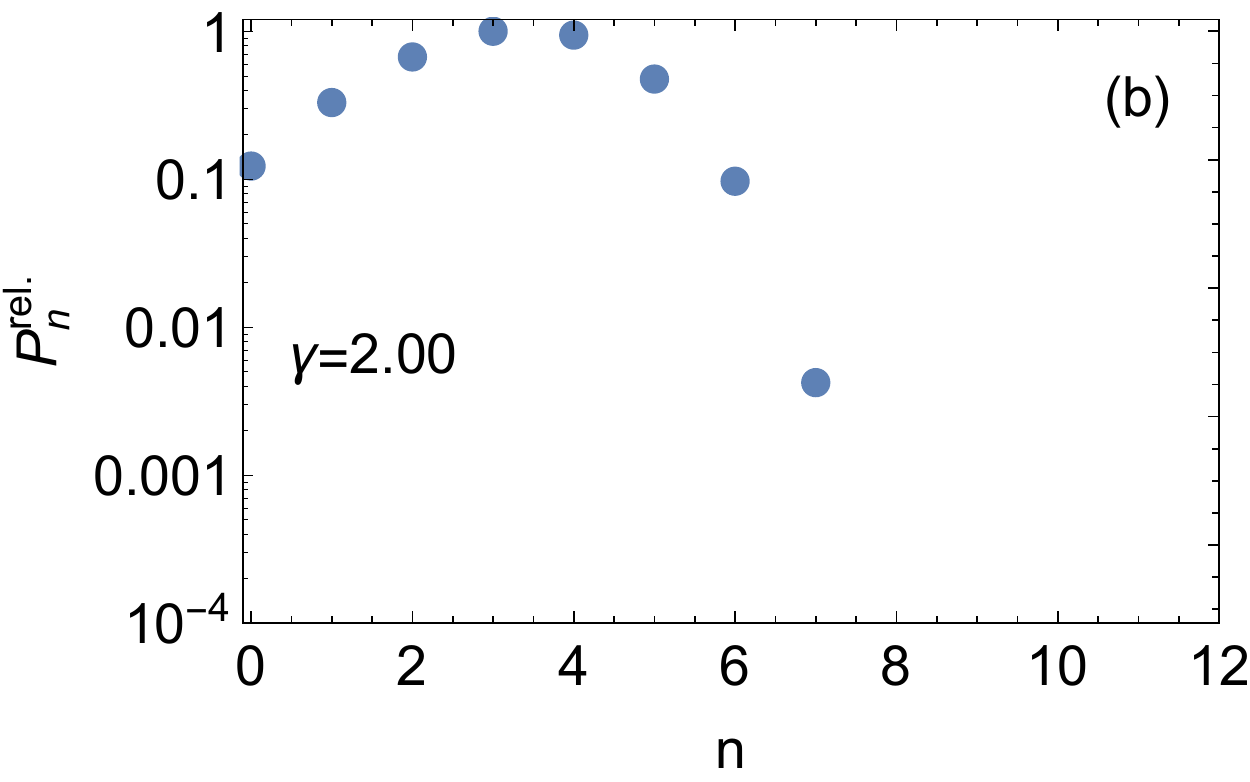}
          \includegraphics[width=0.4\textwidth]{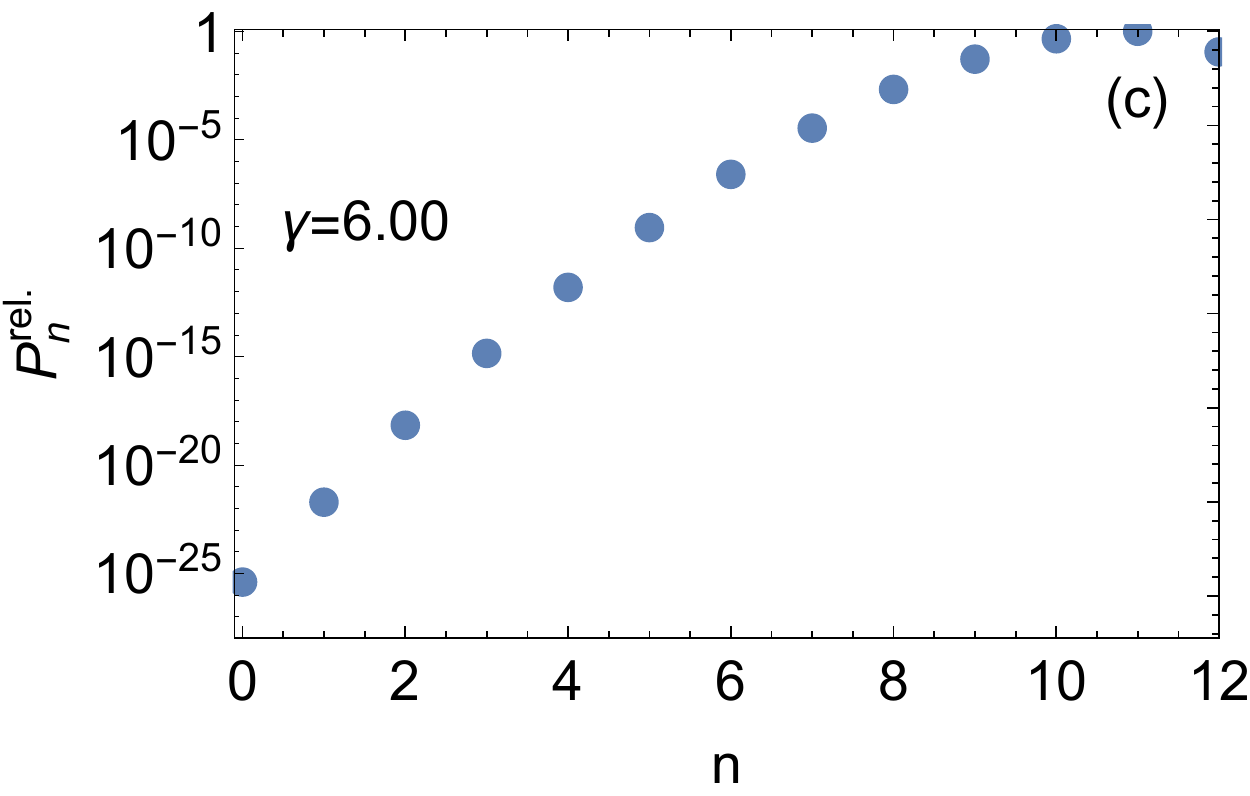}
	   
	\caption{The relative  probabilities of ionization as a function of increasing numbers $n$ of virtually-absorbed photons of frequency $0.025$\, a.u.}
\label{integratedprob}
  \end{center}
  \end{figure}

In Fig.\ref{integratedprob} we show the ionization probabilities as a function of the number of virtually-absorbed photons, up to an energy corresponding to the top of the potential barrier. To illustrate the relative probabilities for ionization from each virtual state, we plot the quantity
\begin{equation}
P_n^{rel.} = \frac{|\Psi^+_n(\vec r_f)|^2}{|\Psi^+_{max}(\vec r_f)|^2}
\end{equation}
for each $n$, where $|\Psi^+_{max}|^2$ corresponds to the $n$ value giving maximum ionization probability and $\mathbf{r}_f$ is the tunnel exit.
In all cases, $\omega$ is fixed at a value $0.025$\,a.u. Direct photo-ionization corresponds to absorption of $20$ photons. For small $\gamma$ equal to $0.66$ , shown in Fig.\ref{integratedprob}a and corresponding to 
a field strength of $0.04$\,a.u., ionization occurs most probably directly out of the ground state. Principally this is because the height and width of the potential barrier falls with increasing maximum field strength and tunnelling probability depends upon it exponentially. Here the barrier is such that tunnelling can take place from the ground state. Although tunnelling from higher energy states is even more probable, this is more than offset by the reduced probability of photon absorption, leading to a monotonic decrease of ionization probability as a function of the number of photons absorbed. 

By contrast, at high value $6.0$ of 
$\gamma$, Fig.\ref{integratedprob}c, the picture is quite different. In this case the field strength is only $ 0.003$a.u. and the height of the barrier suppresses tunnelling strongly from the lower-energy states. Here one sees a monotonic  increase of probability  with photons absorbed corresponding to enhanced tunnelling out of successively higher-energy states. Clearly this limit corresponds to direct multiphoton ionization, as in the ATI calculations of \cite{Kracke}. Paradoxically, it is the increasing tunnelling rate  that leads to the increase of  ionization probability with photon number, what is normally referred to as the multiphoton ionization limit. As field strength decreases this leads in turn to the most probable transition being due to no tunnelling at all, i.e. over-the-barrier release of electrons.

 Note that we are comparing always {\it relative}
probabilities as a function of photons absorbed. Since the field strength is ten times lower, the {\it absolute} probabilities are lower for $\gamma = 6.0$ than for $\gamma= 0.66$ due to the lower field strength for fixed photon frequency.
  
\begin{figure} 
    \begin{center}
      \includegraphics[width=0.4\textwidth]{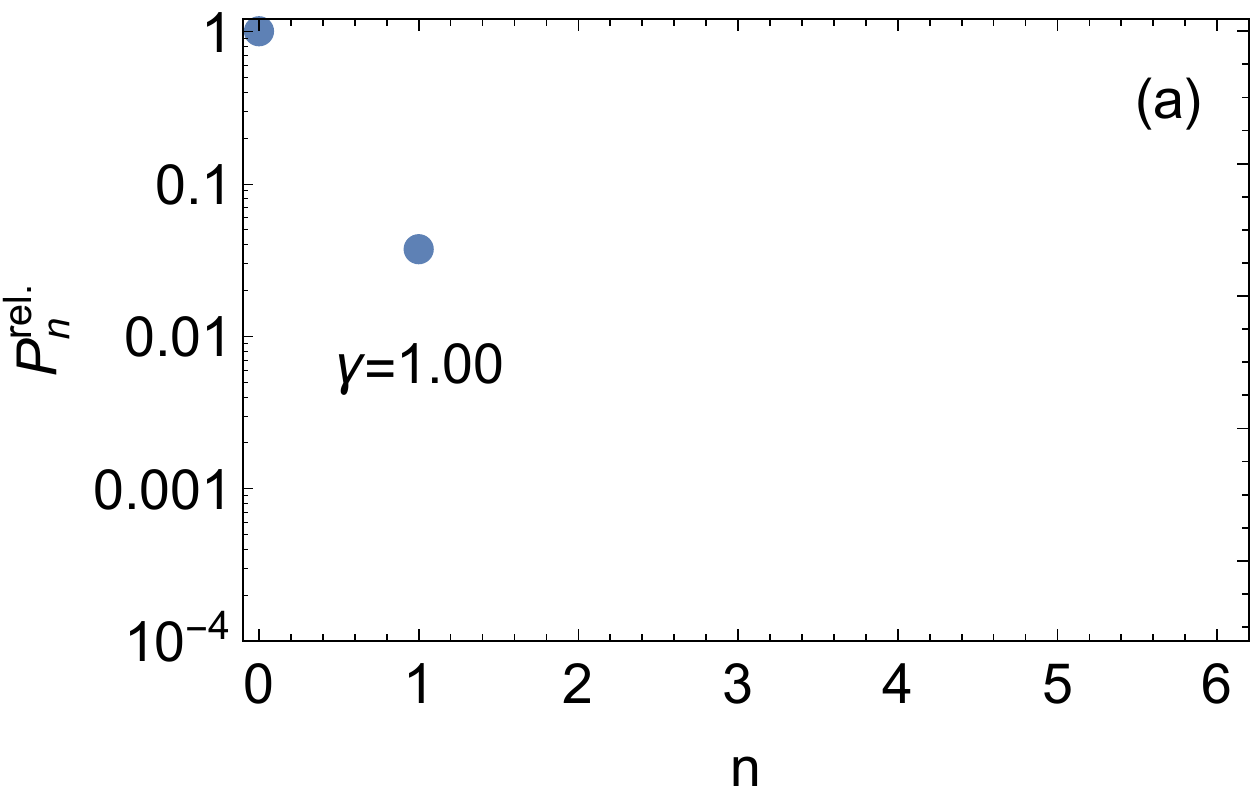}
       \includegraphics[width=0.4\textwidth]{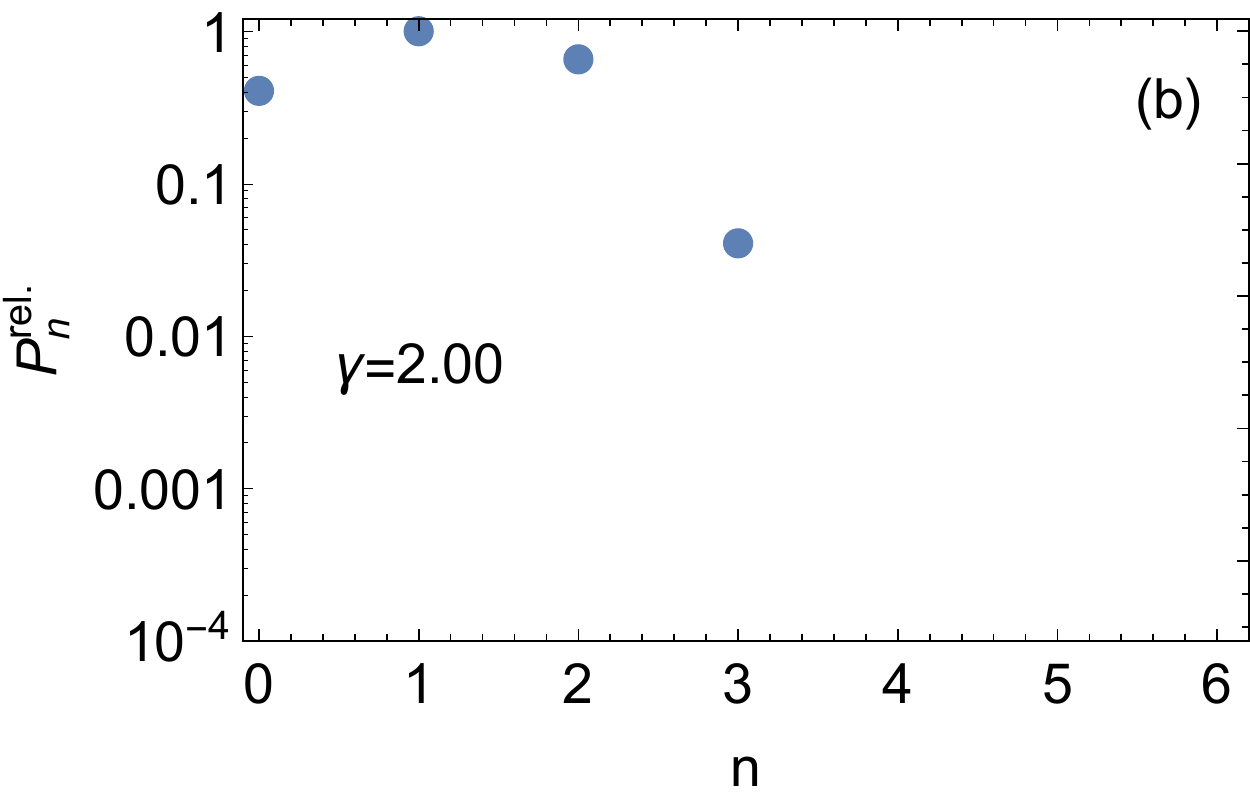}
        \includegraphics[width=0.4\textwidth]{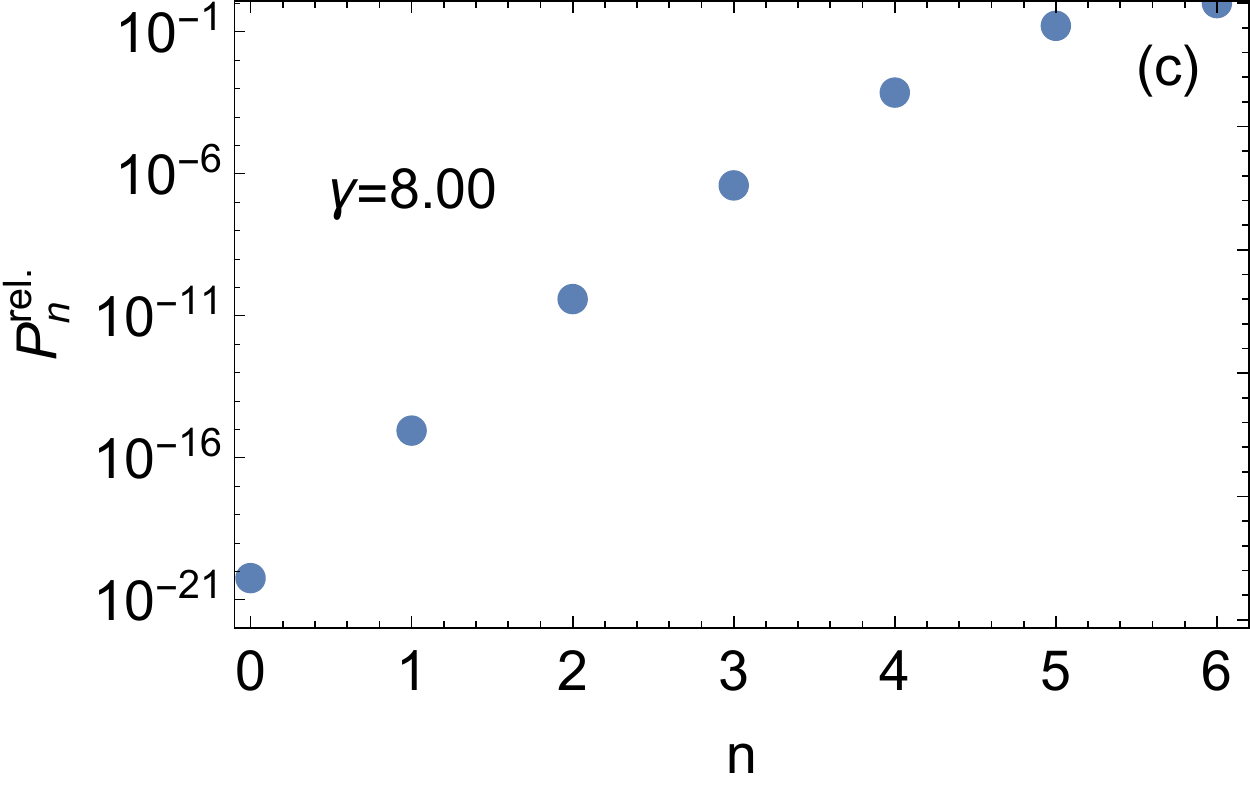}
		\caption{The relative  probabilities of ionization as a function of increasing numbers $n$ of virtually-absorbed photons for $\omega= 0.05$ a.u..}
 \label{4prob005}
  \end{center}
  \end{figure}
  
 As one might expect, intermediate values of $\gamma$ interpolate between these two limits and exhibit preferential tunnelling from virtual excited states which are still below the top of the barrier. For the higher-energy states the probability falls again even though tunnelling is enhanced. The example $\gamma=2.0$, corresponding to field strength $0.01$a.u. is shown in Fig.\ref{integratedprob}b. Here the calculation predicts that a maximum ionization probability  occurs for three photon absorption, followed by tunnelling. The three panels of Fig.\ref{integratedprob}  illustrate nicely the cross-over from ground-state tunnelling to multiphoton ionization without tunnelling.
 
 The same pattern emerges for different values of $\omega$, although for higher frequencies, since for hydrogen $\gamma = \omega/E_0$, the cross-over to predominant ground-state tunnelling (and even over-the-barrier escape) occurs at values of $\gamma$ exceeding unity. This is illustrated in the subsequent figures where we consider the two cases of $\omega = 0.05$ a.u. and $0.1$ a.u..

\begin{figure} 
    \begin{center}
      \includegraphics[width=0.4\textwidth]{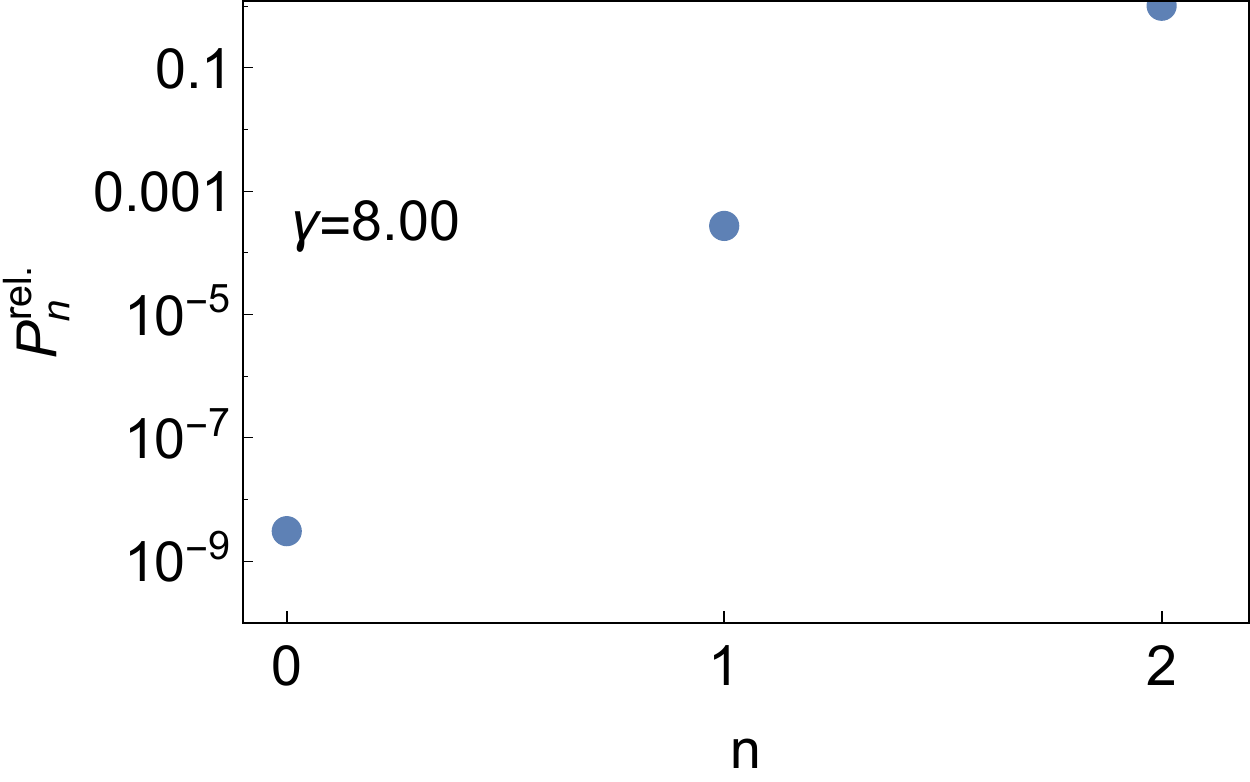}
     		\caption{The relative  probabilities of ionization as a function of increasing numbers $n$ of virtually-absorbed photons for $\omega= 0.10$ a.u..}
 \label{5prob010}
  \end{center}
  \end{figure}

In  Fig.\ref{4prob005}a, we show the relative probabilities for $\omega = 0.05$ a.u. and $\gamma = 1.0$ giving $E_0 = 0.05$ a.u.. Already for this field strength there is most probable tunnelling out of the ground state and the multiphoton excitation probability falls off rapidly for $n = 1$.  At intermediate $\gamma = 2.0$, shown in Fig.\ref{4prob005}b, the most probable tunnelling has shifted to $n = 1$. At higher $\gamma = 8$, Fig.\ref{4prob005}c, corresponding to field strength $E_0 = 0.006$, the transition to preferential multiphoton ionization has been made completely. The same is true for $\omega = 0.10$ and $\gamma=8.0$ shown in Fig.\ref{5prob010}. However, for this frequency, the field is $E_0=0.012$ and only the virtual absorption of two photons is necessary to reach the top of the barrier. In the case of $\omega =0.10$ (not shown), already at $\gamma = 2.0$, the field $E_0 = 0.05$ is such that the barrier is so low that ionization occurs over the barrier after the absorption of just one photon and there is essentially no tunnelling.
  
\section{Conclusions}
We have derived a simple intuitive expression, \eref{poswf} describing the ionization of the hydrogen atom by a classical laser field as consisting of two steps. The ionization amplitude then factors into a product of the separate amplitudes of the two steps occurring. The first step is a polarization of the atom and energy increase of the electron due to the virtual absorption of photons. The second step is the tunnelling of the virtually-excited electron out of the total (atom + field) static potential leading to ionization.

The virtual absorption of photons leads to the electron gaining energy as it recedes from the nucleus and this mechanism supports the supposition of Klaiber et.al. \cite{Klaiber1} who described the ``non-adiabatic" energy gain by classical mechanics. Indeed, there is close agreement, for large number of photons absorbed, between the energy gain predicted in our quantum perturbation theory and that ascribed to classical motion.

The results for the relative ionization probabilities as a function of the number of virtually absorbed photons are presented for fixed laser frequencies but low enough as to
 require many photons to be absorbed to reach the ionization threshold. The Keldysh parameter $\gamma$ then is inversely proportional to the peak field strength and this decides the position and value of the peak of the tunnelling potential that is decisive for the tunnelling probability. The results demonstrate a continuous smooth transition between the two limits in which the maximum probability is associated with direct ground-state tunnelling for the higher strengths and complete multiphoton absorption for lower strength fields.
 
\begin{figure} 
    \begin{center}
      \includegraphics[width=0.4\textwidth]{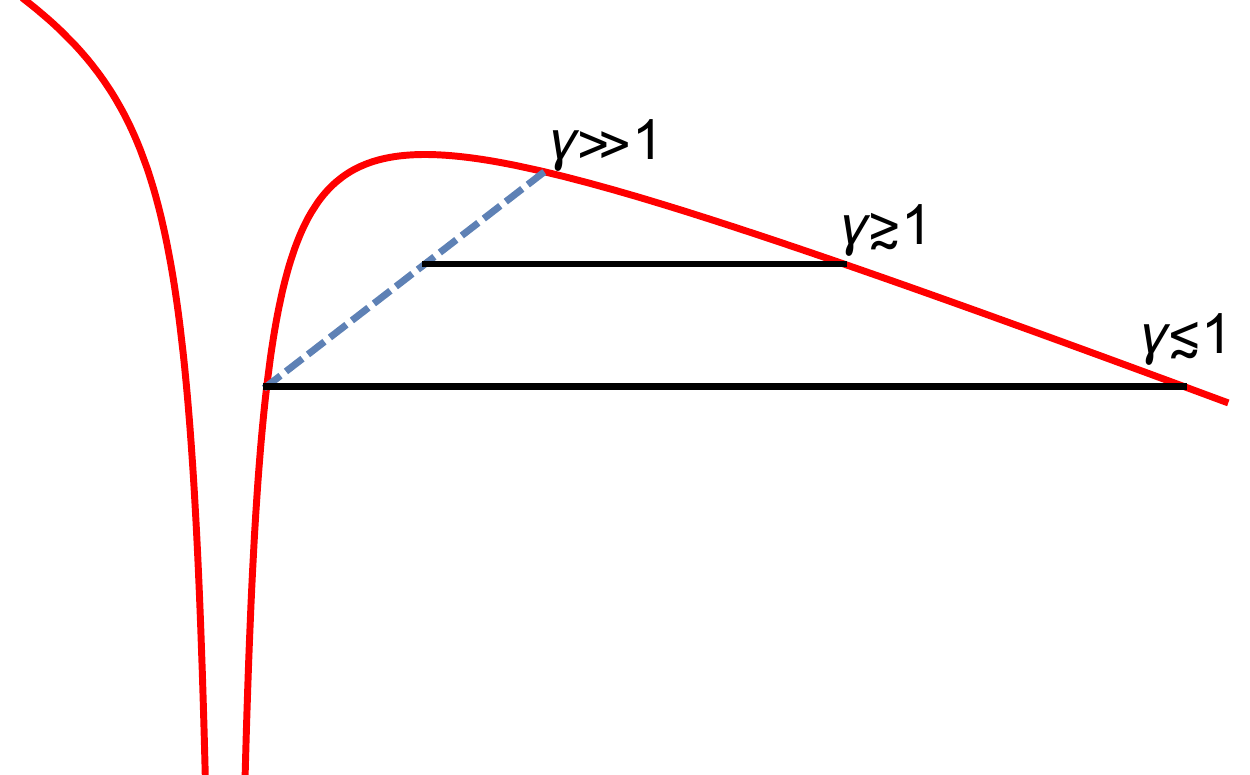}
     		\caption{Schematic picture of the cross-over from tunnelling to multiphoton ionization}
 \label{transition}
  \end{center}
  \end{figure}

 The transition is indicated schematically in Fig.\ref{transition} which emphasizes that, contrary to the usual depiction of a vertical transition in space, as the electron absorbs energy by virtual photon absorption, the atomic wavefunction swells in size. Schematically and following tradition, $\gamma \approx 1$ is shown as the intermediate cross-over region. However, as we have seen in the example of $\omega = 0.10$ a.u., the tunnelling region and indeed direct over-the-barrier field ionization can set in for $\gamma$ values greater than unity.

\section{Acknowledgement}

The authors thank Karen Hatsagortsyan and Christoph Keitel for their comments on the current manuscript.

\begin{appendix}
\section{Off-shell Coulomb wavefunctions}
The inhomogeneous equation for the off-shell Coulomb wavefunction is given in \eref{offshellwfn}
\begin{eqnarray}
\begin{split}
[H_0 + (I_p - n\omega)]\ket{\psi_n} =&  \frac{E_0}{2}(x + iy)\ket{\psi_{n-1}}\nonumber.
\end{split}
\end{eqnarray}

In spherical coordinates this is
\begin{eqnarray}
[H_0 + (I_p - n\omega)]\ket{\psi_n}  = \frac{r E_0}{2}\sin(\theta)~e^{i\phi}|\psi_{n-1}\rangle
\end{eqnarray}

Projecting on spherical harmonics yields an iterative equation for the radial part of the excited states $\psi_n$. With ${\psi}_0(r)=R_0(r)Y_{0,0}(\theta,\phi)/r$ and $R_0(r)=2\sqrt{\kappa^3}r\exp(-\kappa r)$ the first order equation reads:
\begin{eqnarray}
\begin{split}
  R_1''(r) - \frac{2R_1(r)}{r^2} + 
      &\left(2\left(-I_p + \omega + \frac{\kappa}{r}\right)\right) R_1(r)\\& = 
     -\sqrt{\frac{2}{3}} r E_0 R_0(r).
     \end{split}
\end{eqnarray}
The $n$th order equation to be solved iteratively is
\begin{eqnarray}
\label{Aradial}
\begin{split}
  {R}_{n}''(r )+&\left(-\kappa_n^2-\frac{n(n+1)}{r ^2}+\frac{2\kappa}{r}\right) {R}_n(r )\\&= -\sqrt{\frac{2n}{2n+1}}r  E_0  {R}_{n-1}(r ),\\
&\equiv f(r)
\end{split}
\end{eqnarray}
where $\kappa_n^2 \equiv 2(I_p - n\omega)$.

The homogeneous equation has two solution functions that are:
\begin{eqnarray}
y_1(r)&=&W_{\frac{1}{\kappa _n},n+\frac{1}{2}}\left(2 r \kappa _n\right)\nonumber\\
y_2(r)&=&M_{\frac{1}{\kappa _n},n+\frac{1}{2}}\left(2 r \kappa _n\right),
\end{eqnarray}
where $M$ and $W$ are the Whittaker-functions.
Taking into account the asymptotic behavior of these functions the $n$th off-shell wavefunction in the 3D-Coulomb potential can be given via the expression for the radial functions
\begin{eqnarray}
\begin{split}
 R_n(r)= &-y_1(r)\int^r_0 dz\frac{f(z)y_2(z)}{W(z)}\\& +y_2(r)\int^r_{\infty} dz\frac{f(z)y_1(z)}{W(z)}.
 \end{split}
\end{eqnarray}
with the Wronskian $W=y_1y_2'-y_2y_1'$.

 \section{Imaging Theorem}
 
  With a final measured momentum state $\ket{\vec p}$ the probability amplitude $f(t)$, the projection on the exact time-propagating state, can be written $f(t) =  \braket{\vec p}{\Psi^+(t)} \equiv \tilde\Psi^+(\vec p,t)$. One notes that for a free electron this is just the Fourier transform of the exact spatial wavefunction. However, it is defined more generally e.g. in an asymptotic Coulomb potential. 
The imaging theorem (IT) \cite{JBJF} shows that in the region where the semi-classical wavefunction is valid, the amplitude   
 $\tilde\Psi^+(\vec p,t)$ can be related to the position wavefunction $ \Psi^+(\vec r t)$ of \eref{poswf}, which is the quantity we calculate. Specifically, the IT equates the probabilities,
 \begin{equation}
\left|\tilde\Psi^+(\vec p,t)\right|^2~d \vec p = \left|\Psi^+(\vec r,t)\right|^2~d \vec r
\end{equation}
at all points connecting the momentum $\vec p$ with position $\vec r$ (and vice versa) along a  {\it{classical}} trajectory. 
 We have put $\vec r = \vec r_f$, the position corresponding to the barrier exit for each virtual state energy. Thereby we equate the probability density in position space with the probability density for projection 
 onto a momentum state. This means that we have assumed that semi-classics is valid immediately following the electron's transition to a continuum state. This is an approximation but is compatible with our use of semi-classics to describe the under-the-barrier motion. The IT also lends credence to the strategy of Ni et.al. \cite{JMR} who use classical mechanics to propagate numerically-calculated probability densities backwards in time to the tunnelling region.

\end{appendix}


\end{document}